\documentclass{article}%
\usepackage{amsmath}
\usepackage{amsfonts}
\usepackage{amssymb}
\usepackage{graphicx}%
\providecommand{\U}[1]{\protect\rule{.1in}{.1in}}

\begin{document}

\author{Giuseppe Castagnoli
\and Former Elsag Bailey ICT Division, 16031 Pieve Ligure, Italy
\and E-mail: giuseppe.castagnoli@gmail.com}
\title{Origin of the quantum speed-up}
\maketitle

\begin{abstract}
Bob chooses a function from a set of functions and gives Alice the black box
that computes it. Alice is to find a characteristic of the function through
function evaluations. In the quantum case, the number of function evaluations
can be smaller than the minimum classically possible. The fundamental reason
for this violation of a classical limit is not known. We trace it back to a
disambiguation of the principle that measuring an observable determines one of
its eigenvalues. Representing Bob's choice of the label of the function as the
unitary transformation of a random quantum measurement outcome shows that: (i)
finding the characteristic of the function on the part of Alice is a
by-product of reconstructing Bob's choice and (ii) because of the quantum
correlation between choice and reconstruction, one cannot tell whether Bob's
choice is determined by the action of Bob (initial measurement and successive
unitary transformation) or that of Alice (further unitary transformation and
final measurement). Postulating that the determination shares evenly between
the two actions, in a uniform superposition of all the possible ways of
sharing, implies that quantum algorithms are superpositions of histories in
each of which Alice knows in advance one of the possible halves of Bob's
choice. Performing, in each history, only the function evaluations required to
classically reconstruct Bob's choice given the advanced knowledge of half of
it yields the quantum speed-up. In all the cases examined, this goes along
with interleaving function evaluations with non-computational unitary
transformations that each time maximize the amount of information about Bob's
choice acquired by Alice with function evaluation.

\end{abstract}

\section{Executive summary}

By "quantum speed-up" one means the higher efficiency of quantum algorithms
with respect their classical equivalent. Let us provide at once a simple
example of speed-up. Bob hides a ball in one of four drawers, Alice is to
locate it by opening drawers. In the classical case, to be sure of locating
the ball, Alice should plan to open three drawers. With Grover's quantum
database search algorithm $\left[  1\right]  $, only one drawer suffices.

It should be noted that Grover's algorithm, like the seminal one of Deutsch
$\left[  2\right]  $, requires fewer computation steps (drawer openings in
Grover's case) than the minimum demonstrably required by any equivalent
classical algorithm.

As already noted in literature $\left[  3\right]  $, this violation of a limit
applying to any classical time-evolution relates the speed-up to the violation
of the temporal Bell inequality of Leggett and Garg $\left[  4\right]  $, the
information-theoretic one of Braunstein and Caves $\left[  5\right]  $ and,
particularly, the one formulated by Morikoshi $\left[  3\right]  $ exactly in
the case of Grover's algorithm. According to this latter inequality, all is as
if quantum information processing exploited unperformed computations $\left[
3\right]  $. The fundamental reason for this is not known. Here we trace it
back to a disambiguation of the quantum principle -- stating that the
measurement of an observable determines one of its eigenvalues{\normalsize .
As we will see, this principle becomes typically ambiguous in presence of
quantum speed-up.}

\ We focus on quantum oracle computing. Bob chooses a function from a set of
functions and gives Alice the black box (\textit{oracle}) that computes it.
Alice is to find a characteristic of the function chosen by Bob by performing
function evaluations (in Grover's case, opening drawers amounts to evaluating
the Kronecker function).

Our argument goes as follows -- it is clearer to segment it by section.

2 \textit{Grover's algorithm}

We use a representation where Grover's algorithm is the model for all the
quantum algorithms based on function evaluation.

2.1 \textit{Time-symmetric representation}

To the usual \textit{Alice's register}, containing the number of the drawer
that Alice wants to open, we add an imaginary \textit{Bob's register}%
\footnote{We take the expression "imaginary register" from reference $\left[
6\right]  $, which highlights the problem-solution symmetry of Grover's and
the phase estimation algorithms.}, containing the number of the drawer with
the ball. We assume that the initial state of Bob's register is maximally
mixed, so that Bob's process of choice is represented from scratch. See the
far left of Fig. 1, where S. stands for state, M. for measurement, $U$\ is the
unitary part of the quantum algorithm.%

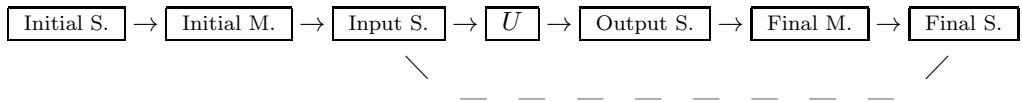
\begin{align*}%
\begin{tabular}
[c]{|c|}\hline
{\footnotesize Initial S.}\\\hline
\end{tabular}
&  \rightarrow%
\begin{tabular}
[c]{|c|}\hline
{\footnotesize Initial M.}\\\hline
\end{tabular}
\rightarrow%
\begin{tabular}
[c]{|c|}\hline
{\footnotesize Input S.}\\\hline
\end{tabular}
\rightarrow%
\begin{tabular}
[c]{|c|}\hline
${\footnotesize U}$\\\hline
\end{tabular}
\rightarrow%
\begin{tabular}
[c]{|c|}\hline
{\footnotesize Output S.}\\\hline
\end{tabular}
\rightarrow%
\begin{tabular}
[c]{|c|}\hline
{\footnotesize Final M.}\\\hline
\end{tabular}
\rightarrow%
\begin{tabular}
[c]{|c|}\hline
{\footnotesize Final S.}\\\hline
\end{tabular}
\\
&  \ ~~~~~~~~~~~\ \ \ \ \ \ \ \ \ \ \ \ \ \ \ \ \ \
\begin{tabular}
[c]{cccccccccc}%
$\diagdown$ &  &  &  &  &  &  &  &  & $\diagup$\\
& --- & --- & --- & --- & --- & --- & --- & --- &
\end{tabular}
\\
&  ~\ \ \ \ \ \ \text{{\footnotesize Quantum correlation, reading the output
contributes to determining the input}}\\
&  \text{Fig. 1}\ \text{Time-symmetric representation of quantum algorithms }%
\end{align*}

Bob measures the content of this register obtaining a drawer number uniformly
at random. To start with, we assume that Bob's choice is this very number. The
corresponding eigenstate, with the usual sharp state of Alice's register, is
the input of $U$. Here, by performing function evaluations (by opening
drawers), Alice reconstructs Bob's choice in her register. By finally
measuring the content of this register, she acquires the number of the drawer
chosen by Bob.

In this extended representation of the quantum algorithm there is quantum
correlation between the contents of Bob's and Alice's registers before their
respective measurements. These in fact yield two identical eigenvalues whose
common value (the number of the drawer chosen by Bob) is selected at random.
We will see that quantum correlation remains there also when Bob unitarily
changes the initial random measurement outcome into a desired number.

By \textit{time-symmetric} $\left[  7,8\right]  $\ representation of the
quantum algorithm we mean the present representation (extended to Bob's
choice), with the peculiarity that the projection of the quantum state due to
Bob's measurement is retarded to the end of $U$. As well known, such
projections can be retarded or advanced along a unitary transformation that
follows or precedes the measurement. In the present case, retarding the
projection \textit{relativizes} the quantum state to the observer Alice in the
sense of relational quantum mechanics $\left[  9\right]  $. Alice is in fact
forbidden to observe the result of Bob's measurement before reconstructing it
through function evaluations. In this relativized representation, the
maximally mixed initial state of Bob's register remains unaltered after Bob's
measurement. Its entropy represents Alice's ignorance of Bob's choice.

2.2 \textit{Sharing the determination of Bob's choice}

The quantum principle, stating that the measurement of an observable
determines one of its eigenvalues, becomes ambiguous when the measurement of
two commuting observables yields at random two identical eigenvalues (choice
and reconstruction). Which measurement determines their common value? The idea
that all the determination should be ascribed to the measurement performed
first is not justified. In fact Bob's measurement can be suppressed and the
determination of Bob's choice is performed by Alice's measurement, also at the
time of the suppressed measurement -- the projection of the quantum state due
to Alice's measurement (i. e. the determination) can be advanced at the time
in question by applying $U^{\dag}$ to the two ends of it. Since there is no
way of telling which measurement determines Bob's choice, for reasons of
symmetry we postulate that the determination shares between the two
measurements (i) without over-determination (i. e. without producing twice the
same information), (ii)\ with entropy reductions the same for each share, and
(iii) in a uniform quantum superposition of all the possible ways of sharing
compatible with the former conditions. Conditions (i) and (ii) imply that
Alice's measurement determines half of Bob's choice ($n/2$ bits in the present
case where Bob's choice is an unstructured $n$\ bit string). For condition
(iii), the quantum algorithm should be seen as a uniform quantum superposition
of algorithms (histories) in each of which Alice's measurement determines one
of the possible halves of Bob's choice. We call conditions (i) through (iii)
the\textit{ sharing rule}. This rule has been inspired by the work of Dolev
and Elitzur $\left[  10\right]  $ on the non-sequential behavior of the wave
function highlighted by partial measurement. Here partial measurements are
involved in sharing the determination of Bob's choice.

2.3 \textit{Advanced knowledge}

\textit{ }By advancing (by $U^{\dag}$) to the beginning of Alice's action
(immediately after Bob's measurement) the contribution of Alice's measurement
to the determination of Bob's choice, the maximally mixed initial state of
Bob's register is projected on a less mixed state where the corresponding half
of Bob's choice is determined. Correspondingly, the entropy of the state is
halved. This means that, in each history, Alice knows half of Bob's choice in advance.

2.4 \textit{The mechanism of the speed-up in Grover's algorithm}

According to the sharing rule (the present disambiguation of the quantum
principle), the quantum algorithm is a superposition of histories in each of
which Alice knows in advance one of the possible halves of Bob's choice. It
should be noted that this holds for any quantum algorithm that reconstructs
Bob's choice, with of without speed-up. The quantum correlation between choice
and reconstruction is anyhow there. Thus, at one extreme, the quantum
algorithm can be a superposition of identical histories in each of which Alice
ignores the advanced knowledge that tags the history and performs the function
evaluations classically required to reconstruct Bob's choice. At the other, in
each history, Alice should be able to perform only the function evaluations
required to classically identify the missing half of Bob's choice given the
advanced knowledge of the other half; in fact, this is what is needed to bring
the halved entropy of Bob's register down to zero. In Grover's algorithm, this
is made possible by interleaving function evaluations with non-computational
unitary transformations applying to Alice's register that each time maximize
the amount of information about Bob's choice acquired by Alice with function
evaluation. This minimizes the number of function evaluations bringing it
exactly to the number ($\mathcal{N}_{a}$) required to reconstruct Bob's choice
given the advanced knowledge of half of it. This explains why Grover's
algorithm requires $\mathcal{N}_{a}=\operatorname{O}\left(  2^{n/2}\right)  $
function evaluations against the\ $\operatorname{O}\left(  2^{n}\right)  $\ of
the classical case and why the violation of Morikoshi's inequality implies
that it exploits unperformed computations. This is what happens in each and
every history the algorithm is made of.

3 \textit{Generalizing the mechanism of the speed-up}

A simple generalization of Grover's algorithm produces all the quantum
algorithms whose solution is a by-product of the reconstruction of Bob's
choice. First, we should set the non-computational unitary transformations
free. Then we should determine them by maximizing each time, after the
transformation that follows function evaluation, the probability of finding
the solution in Alice's register. This minimizes the number of function
evaluations, bringing it to $\mathcal{N}_{a}$ in all the cases examined. Given
the set of functions, this mechanism produces the quantum algorithm that
yields the solution (the characteristic of the function chosen by Bob) with
the maximum possible speed-up.

4 \textit{Deutsch\&Jozsa's algorithm}, 5 \textit{Simon's and the hidden
subgroup algorithms}

Here Bob's choice is a highly structured bit string. Given the advanced
knowledge of half of it according to the sharing rule, finding the missing
half requires a single function evaluation -- against an exponential number
thereof in the absence of advanced knowledge. This explains the exponential
speed-up of these latter algorithms.

6 \textit{Discussion and conclusions}

We have identified the fundamental reason for which some quantum algorithms
violate a limit applying to classical time-evolutions and/or Morikoshi's
inequality. Although preliminary in character, the results obtained seem to
open a gap in a problem that has remained little explored. Until now there was
no fundamental explanation of the speed-up, no general mechanism for producing it.

With respect to references $\left[  11,12\right]  $, we have reformulated the
explanation of the speed-up given for Grover's algorithm and extended it to
all the quantum algorithms based on function evaluation.

\section{Grover's algorithm}

We develop our argument in detail for Grover's algorithm. Its time-symmetric
representation is the model for all the quantum algorithms examined in this paper.

\subsection{Time-symmetric representation}

Let $\mathbf{b}$ and $\mathbf{a}$, ranging over $\left\{  0,1\right\}  ^{n}$,
be respectively the number of the drawer with the ball and that of the drawer
that Alice wants to open. Bob writes his choice of the value of $\mathbf{b}$
in an imaginary $n$-qubit register $B$. Alice writes a value of $\mathbf{a}$
in a $n$-qubit register $A$. Then the black box computes the Kronecker
function $\delta\left(  \mathbf{b},\mathbf{a}\right)  $, which gives $1$ if
$\mathbf{b}=\mathbf{a}$ and $0$\ otherwise -- tells Alice whether the ball is
in drawer $\mathbf{a}$. A one-qubit register $V$ is meant to contain the
result of the computation of $\delta\left(  \mathbf{b},\mathbf{a}\right)  $ --
modulo 2 added to its former content for logical reversibility.

We assume that register $B$ is initially in a maximally mixed state, so that
the value of $\mathbf{b}$\ is completely undetermined. We will see that this
assumption just yields a special view of the usual quantum algorithm (starting
with a completely determined value of $\mathbf{b}$\ ). Registers $A$ and
$V$\ are prepared as usual in a sharp state. With $n=2$, the initial state of
the three registers is thus:%

\begin{equation}
\left\vert \psi\right\rangle =\frac{1}{2}\left(  \operatorname{e}%
^{i\varphi_{0}}\left\vert 00\right\rangle _{B}+\operatorname{e}^{i\varphi_{1}%
}\left\vert 01\right\rangle _{B}+\operatorname{e}^{i\varphi_{2}}\left\vert
10\right\rangle _{B}+\operatorname{e}^{i\varphi_{3}}\left\vert 11\right\rangle
_{B}\right)  \left\vert 00\right\rangle _{A}\left\vert 1\right\rangle _{V}.
\label{init}%
\end{equation}
We keep the usual state vector representation of quantum algorithms by using
the random phase representation of density operators $\left[  13\right]  $.
The $\varphi_{i}$ are independent random phases each with uniform distribution
in $\left[  0,2\pi\right]  $. The density operator is the average over all
$\varphi_{i}$ of the product of the ket by the bra:%
\begin{align*}
\left\langle \left\vert \psi\right\rangle \left\langle \psi\right\vert
\right\rangle _{\forall\varphi_{i}}  &  =\frac{1}{4}\left(  \left\vert
00\right\rangle _{B}\left\langle 00\right\vert _{B}+\left\vert 01\right\rangle
_{B}\left\langle 01\right\vert _{B}+\left\vert 10\right\rangle _{B}%
\left\langle 10\right\vert _{B}+\left\vert 11\right\rangle _{B}\left\langle
11\right\vert _{B}\right) \\
&  \left\vert 00\right\rangle _{A}\left\langle 00\right\vert _{A}\left\vert
1\right\rangle _{V}\left\langle 1\right\vert _{V}.
\end{align*}
\ The von Neumann entropy of the state of register $B$ in the overall state
(\ref{init}) is two bits. This is also the entropy of the overall quantum
state. As we will see, this latter entropy coincides with that of the reduced
density operator of register $B$ throughout the quantum algorithm.

We call $\hat{B}$ ($\hat{A}$) the content of register $B$ ($A$), of eigenvalue
$\mathbf{b}$ ($\mathbf{a}$). $\hat{B}$ and $\hat{A}$, both diagonal in the
computational basis, commute. To prepare register $B$ in the desired value of
$\mathbf{b}$, in the first place Bob should measure $\hat{B}$\ in state
(\ref{init}). He obtains an eigenvalue at random, say $\mathbf{b}=01$.
Conventionally, state (\ref{init}) would be projected on:%
\begin{equation}
P_{B}\left\vert \psi\right\rangle =\left\vert 01\right\rangle _{B}\left\vert
00\right\rangle _{A}\left\vert 1\right\rangle _{V}. \label{initb}%
\end{equation}
For the time being, we assume that Bob's choice is random, is the result of
measurement itself. The case that Bob chooses a predetermined value of
$\mathbf{b}$ is considered further on$.$

State (\ref{initb}), with register $B$\ in a sharp state, is the input state
of the conventional representation of the quantum algorithm. For reasons that
will become clear, we retard to the end of the unitary part of the algorithm
the projection of state (\ref{init}) on state (\ref{initb}). Thus, the input
state of the algorithm is state (\ref{init}) back again.

At this point, Alice applies the Hadamard transforms $U_{A}$\ and $U_{V}$ to
respectively registers $A$ and $V$:%
\begin{align}
U_{A}U_{V}\left\vert \psi\right\rangle  &  =\frac{1}{4\sqrt{2}}\left(
\operatorname{e}^{i\varphi_{0}}\left\vert 00\right\rangle _{B}%
+\operatorname{e}^{i\varphi_{1}}\left\vert 01\right\rangle _{B}%
+\operatorname{e}^{i\varphi_{2}}\left\vert 10\right\rangle _{B}%
+\operatorname{e}^{i\varphi_{3}}\left\vert 11\right\rangle _{B}\right)
\nonumber\\
&  \left(  \left\vert 00\right\rangle _{A}+\left\vert 01\right\rangle
_{A}+\left\vert 10\right\rangle _{A}+\left\vert 11\right\rangle _{A}\right)
\left(  \left\vert 0\right\rangle _{V}-\left\vert 1\right\rangle _{V}\right)
. \label{tred}%
\end{align}

Then she performs the reversible computation of $\delta\left(  \mathbf{b}%
,\mathbf{a}\right)  $, represented by the unitary transformation $U_{f}$ ($f$
like "function evaluation"):%
\begin{equation}
U_{f}U_{A}U_{V}\left\vert \psi\right\rangle =\frac{1}{4\sqrt{2}}\left[
\begin{array}
[c]{c}%
\operatorname{e}^{i\varphi_{0}}\left\vert 00\right\rangle _{B}\left(
-\left\vert 00\right\rangle _{A}+\left\vert 01\right\rangle _{A}+\left\vert
10\right\rangle _{A}+\left\vert 11\right\rangle _{A}\right)  +\\
\operatorname{e}^{i\varphi_{1}}\left\vert 01\right\rangle _{B}\left(
\left\vert 00\right\rangle _{A}-\left\vert 01\right\rangle _{A}+\left\vert
10\right\rangle _{A}+\left\vert 11\right\rangle _{A}\right)  +\\
\operatorname{e}^{i\varphi_{2}}\left\vert 10\right\rangle _{B}\left(
\left\vert 00\right\rangle _{A}+\left\vert 01\right\rangle _{A}-\left\vert
10\right\rangle _{A}+\left\vert 11\right\rangle _{A}\right)  +\\
\operatorname{e}^{i\varphi_{3}}\left\vert 11\right\rangle _{B}\left(
\left\vert 00\right\rangle _{A}+\left\vert 01\right\rangle _{A}+\left\vert
10\right\rangle _{A}-\left\vert 11\right\rangle _{A}\right)
\end{array}
\right]  (\left\vert 0\right\rangle _{V}-\left\vert 1\right\rangle _{V}).
\label{secondstaged}%
\end{equation}

$U_{f}$ maximally entangles registers $B$ and $A$ (i. e. the observables
$\hat{B}$ and $\hat{A}$). Four orthogonal states of $B$, each a value of
$\mathbf{b}$, one by one multiply four orthogonal states of $A$. This means
that the information about the value of $\mathbf{b}$ has propagated to
register $A$.

If we measured $\hat{A}$\ in state (\ref{secondstaged}), we would obtain a
value of $\mathbf{a}$\ completely uncorrelated with that of $\mathbf{b}$. To
make the information acquired with function evaluation accessible to
measurement, we need to make correlation of entanglement. This is done by
applying to register $A$\ the unitary transformation\ $U_{A}^{\prime}$ (the so
called \textit{inversion about the mean}):%
\begin{align}
U_{A}^{\prime}U_{f}U_{A}U_{V}\left\vert \psi\right\rangle  &  =\frac{1}%
{2\sqrt{2}}\left(  \operatorname{e}^{i\varphi_{0}}\left\vert 00\right\rangle
_{B}\left\vert 00\right\rangle _{A}+\operatorname{e}^{i\varphi_{1}}\left\vert
01\right\rangle _{B}\left\vert 01\right\rangle _{A}+\operatorname{e}%
^{i\varphi_{2}}\left\vert 10\right\rangle _{B}\left\vert 10\right\rangle
_{A}+\operatorname{e}^{i\varphi_{3}}\left\vert 11\right\rangle _{B}\left\vert
11\right\rangle _{A}\right) \nonumber\\
&  \left(  \left\vert 0\right\rangle _{V}-\left\vert 1\right\rangle
_{V}\right)  . \label{threed}%
\end{align}
Now the contents of registers $B$ and $A$\ are identical: Alice has
reconstructed Bob's choice in register $A$. She acquires the reconstruction by
measuring $\hat{A}$. This projects state (\ref{threed}) on:%
\begin{equation}
P_{A}U_{A}^{\prime}U_{f}U_{A}U_{V}\left\vert \psi\right\rangle =\frac{1}%
{\sqrt{2}}\left\vert 01\right\rangle _{B}\left\vert 01\right\rangle
_{A}\left(  \left\vert 0\right\rangle _{V}-\left\vert 1\right\rangle
_{V}\right)  , \label{quattrod}%
\end{equation}
in overlap with the retarded projection due to the measurement of $\hat{B}%
$\ in state (\ref{init}). The two projections are redundant with one another.

We call equations (\ref{init}) and (\ref{tred}) through (\ref{quattrod}) the
\textit{time-symmetric representation }of the quantum algorithm. It should be
noted that this representation is the conventional one, starting with a well
determined value of $\mathbf{b}$, relativized to the observer Alice in the
sense of relational quantum mechanics $\left[  9\right]  $. By definition, the
projection due to measuring $\hat{B}$\ in state (\ref{init}) should remain
hidden to the observer Alice until she has reconstructed Bob's choice. It
should in fact be retarded until Alice measures $\hat{A}$ in state
(\ref{threed}).

In this representation, the two bit entropy of state (\ref{init}) represents
Alice's ignorance of Bob's choice. When Alice measures $\hat{A}$ in state
(\ref{threed}), the entropy of the quantum state becomes zero and she acquires
full knowledge of Bob's choice. Thus, the entropy of the quantum state -- or
identically that of the reduced density operator of register $B$ -- gauges
Alice's ignorance of Bob's choice.

We can see that there is quantum correlation between the outcome of measuring
$\hat{B}$\ in state (\ref{init}) and that of measuring $\hat{A}$\ in state
(\ref{threed}). In fact one obtains uniformly at random two identical
eigenvalues, namely Bob's choice -- in present assumptions the value $01$ of
both $\mathbf{b}$ and $\mathbf{a}$. This quantum correlation plays a crucial
role in the present explanation of the speed-up.

Until now we have assumed that Bob's choice is a random quantum measurement
outcome. An equally crucial point of our argument is noting that quantum
correlation remains there also when Bob chooses a predetermined value of
$\mathbf{b}$. Say that the measurement of $\hat{B}$ in state (\ref{init})
yields $\mathbf{b}=11$ and Bob wants $\mathbf{b}=01$. He applies to register
$B$ a permutation of the values of $\mathbf{b}$, a unitary transformation
$U_{B}$ such that $U_{B}\left\vert 11\right\rangle _{B}=\left\vert
01\right\rangle _{B}$. The correlation\ is the same as before up to\ $U_{B}$.
The point is that, from the standpoint of quantum correlation, $U_{B}$\ should
be considered a "fixed" transformation.

In fact quantum correlation concerns two measurement outcomes in an ensemble
of repetitions of the same experiment, consisting of the measurement of an
observable in an initial state, a unitary transformation, and the measurement
of another observable in the resulting state. Initial state and unitary
transformation should remain unaltered throughout the ensemble of repetitions.
$U_{B}$, being part of the unitary transformation, should be considered always
the same.

Thus, from the standpoint of quantum correlation, the predetermined value of
$\mathbf{b}$, seen as the fixed permutation of a random measurement outcome,
should be considered a random measurement outcome as well.

\subsection{Sharing the determination of Bob's choice}

We share the determination of Bob's choice between Bob's and Alice's
measurements or, more exhaustively, \textit{actions}. In fact Bob's choice is
determined by either Bob's measurement and his successive unitary action (to
change the random outcome into the one desired) or Alice's action of unitarily
reconstructing Bob's choice and finally measuring the reconstruction.

First, we introduce the tools required to perform the sharing.

We call $\left\vert \psi\right\rangle _{B}$ the state of register $B$ in the
overall states (\ref{init}) and (\ref{tred}) through (\ref{threed}):
\begin{equation}
\left\vert \psi\right\rangle _{B}=\frac{1}{2}\left(  \operatorname{e}%
^{i\varphi_{0}}\left\vert 00\right\rangle _{B}+\operatorname{e}^{i\varphi_{1}%
}\left\vert 01\right\rangle _{B}+\operatorname{e}^{i\varphi_{2}}\left\vert
10\right\rangle _{B}+\operatorname{e}^{i\varphi_{3}}\left\vert 11\right\rangle
_{B}\right)  . \label{rosuba}%
\end{equation}
$\left\vert \psi\right\rangle _{B}$ is the random phase representation of the
reduced density operator of register $B$:%
\[
\left\langle \left\vert \psi\right\rangle _{B}\left\langle \psi\right\vert
_{B}\right\rangle _{\forall\varphi_{i}}=\frac{1}{4}\left(  \left\vert
00\right\rangle _{B}\left\langle 00\right\vert _{B}+\left\vert 01\right\rangle
_{B}\left\langle 01\right\vert _{B}+\left\vert 10\right\rangle _{B}%
\left\langle 10\right\vert _{B}+\left\vert 11\right\rangle _{B}\left\langle
11\right\vert _{B}\right)  .
\]
It should be noted that the unitary part of Alice's action is the identity on
$\left\vert \psi\right\rangle _{B}$ (it does not change Bob's choice).
$\mathcal{E}_{B}$, the entropy of $\left\vert \psi\right\rangle _{B}$, is two
bits. The determination of Bob's choice is represented by $P_{B}$, the
projection of $\left\vert \psi\right\rangle _{B}$ on $\left\vert
01\right\rangle _{B}$ due to the measurement of either $\hat{B}$ in state
(\ref{init}) or $\hat{A}$ in state (\ref{threed}). We share the determination
of Bob's choice by sharing $P_{B}$, what can be done by resorting to the
notion of partial measurement.

Let us resolve $\mathbf{b}$ into its individual bits: $\mathbf{b}\equiv
b_{0}b_{1}$.\ We consider the following partial measurements and the
corresponding projections of $\left\vert \psi\right\rangle _{B}$. The
measurement of the content of the left cell of register $B$ -- of the
observable $\hat{B}_{0}$ of eigenvalue $b_{0}$ (from now on we omit speaking
of the corresponding operation on register $A$ at the end of the algorithm,
which is completely redundant). A-priori, the measurement outcome is either
$b_{0}=0$ or $b_{0}=1$. However, in present assumptions, the measurement of
$\hat{B}$ projects $\left\vert \psi\right\rangle _{B}$ on $\left\vert
01\right\rangle _{B}$, we are in fact discussing how to share this projection.
Thus we should assume that the measurement of $\hat{B}_{0}$ yields $b_{0}=0$,
namely projects $\left\vert \psi\right\rangle _{B}$ on $\frac{1}{\sqrt{2}%
}\left(  \operatorname{e}^{i\varphi_{0}}\left\vert 00\right\rangle
_{B}+\operatorname{e}^{i\varphi_{1}}\left\vert 01\right\rangle _{B}\right)  $;
we also say "on $\mathbf{b}\in\left\{  01,00\right\}  $". Similarly, the
measurement of the content of the right cell of register $B\ $projects
$\left\vert \psi\right\rangle _{B}$ on $\mathbf{b}\in\left\{  01,11\right\}
$, that of the exclusive or of the contents of the two cells projects
$\left\vert \psi\right\rangle _{B}$ on $\mathbf{b}\in\left\{  01,10\right\}  $.

We will see afterwards that $P_{B}$ should be shared into any two of the three
projections of $\left\vert \psi\right\rangle _{B}$ on: $\mathbf{b}\in\left\{
01,00\right\}  $, $\mathbf{b}\in\left\{  01,11\right\}  $, and $\mathbf{b}%
\in\left\{  01,10\right\}  $. One share (either one) should be ascribed to the
action of Bob, the other to that of Alice.

Until now we have introduced the tools to share the determination of Bob's
choice. Now we introduce some conditions that, reasonably, should be satisfied
by the sharing.

First, we get rid of all redundancy between the two measurements. We resort to
Occam's razor; in Newton's formulation, it states \textquotedblleft\textit{We
are to admit no more causes of natural things than such that are both true and
sufficient to explain their appearances}\textquotedblright\ $\left[
14\right]  $. This requires that, together, the two shares of $P_{B}$\ (the
corresponding partial measurements)\ \textit{tightly} determine the value of
$\mathbf{b}$, namely without determining twice any Boolean function of
$\mathbf{b}$. This is condition (i) of the \textit{sharing rule}.

We apply it to Grover's algorithm. Here, the $n$ bits that specify the value
of $\mathbf{b}$ are independently selected in a random way. Thus, condition
(i) requires that the determination of $p$ of these bits ($0\leq p\leq n$) is
ascribed to the action of Bob, that of the other $n-p$ bits to that of Alice.

Condition (i) does not constrain the value of $p$. This is up to the following
condition (ii). Let $\Delta\mathcal{E}_{B}^{\left(  B\right)  }$ (
$\Delta\mathcal{E}_{B}^{\left(  A\right)  }$)\ be the reduction of the entropy
of the state of register $B$\ associated with the share of $P_{B}$ ascribed to
Bob's (Alice's) action. Here we have $\Delta\mathcal{E}_{B}^{\left(  B\right)
}=p$ bit, $\Delta\mathcal{E}_{B}^{\left(  A\right)  }=\left(  n-p\right)  $
bit. Since Bob's choice is indistinguishably determined by either Bob's or
Alice's action, for reasons of symmetry we require:
\begin{equation}
\Delta\mathcal{E}_{B}^{\left(  B\right)  }=\Delta\mathcal{E}_{B}^{\left(
A\right)  }. \label{equal}%
\end{equation}
Here this becomes $p=n-p=n/2$ -- the\ $n$ bits of $\mathcal{E}_{B}$ share
evenly between the two actions.

We can see that sharing $P_{B}$ into any two of the above said three
projections satisfies conditions (i) and (ii). Any pair of projections,
corresponding to the measurement of a pair of observables among $\hat{B}_{0}$,
$\hat{B}_{1}$, and $\hat{B}_{X}$, tightly selects a value of $\mathbf{b}$. Any
projection reduces the entropy of the state of register $B$ by one bit, so
that equation (\ref{equal}) is always satisfied. We can also see that there is
no other way of satisfying the sharing rule.

Sharing between Bob's and Alice's actions the determination of Bob's choice is
equivalent to saying that Alice's action contributes to this determination.
Thus, in Grover's algorithm, Alice's action determines half of the bits that
specify Bob's choice.

This faces us with the problem that half of Bob's choice can be taken in many
ways. A natural way of solving this problem is requiring that the sharing is
done in a uniform quantum superposition of all the possible ways of taking
half of the choice. This is condition (iii) of the sharing rule. It implies
seeing the quantum algorithm as a uniform superposition of algorithms (or
"histories"), in each of which Alice determines one of the possible halves of
Bob's choice.

\subsection{Advanced knowledge}

We show that ascribing to Alice's action the determination of part of Bob's
choice\ implies that Alice knows in advance, before running the algorithm,
that part of the choice.

For example, we ascribe to Alice's action the determination $b_{0}=0$, namely
the projection of state (\ref{threed}) on%
\begin{equation}
\frac{1}{2}\left(  \operatorname{e}^{i\varphi_{0}}\left\vert 00\right\rangle
_{B}\left\vert 00\right\rangle _{A}+\operatorname{e}^{i\varphi_{1}}\left\vert
01\right\rangle _{B}\left\vert 01\right\rangle _{A}\right)  \left(  \left\vert
0\right\rangle _{V}-\left\vert 1\right\rangle _{V}\right)  . \label{pro}%
\end{equation}
We advance this projection to the beginning of Alice's action, immediately
after Bob's measurement. This is done by applying $U_{V}^{\dag}U_{A}^{\dag
}U_{f}^{\dag}U_{A}^{\prime\dag}$ to the two ends of it, namely to states
(\ref{threed}) and (\ref{pro}). This yields the projection of the input state
of the quantum algorithm (\ref{init}) on%
\begin{equation}
\frac{1}{\sqrt{2}}\left(  \operatorname{e}^{i\varphi_{0}}\left\vert
00\right\rangle _{B}+\operatorname{e}^{i\varphi_{1}}\left\vert 01\right\rangle
_{B}\right)  \left\vert 00\right\rangle _{A}\left\vert 1\right\rangle _{V}.
\end{equation}
Thus, the entropy of the state of register $B$ (or identically of the overall
quantum state) in the input state of the quantum algorithm is halved. Since
this entropy represents Alice's ignorance of Bob's choice (Section 2.1), this
means that Alice, before running the algorithm, knows $n/2$ \ of the bits that
specify Bob's choice, here one bit -- in fact $b_{0}=0$.

According to the sharing rule, the quantum algorithm is a superposition of
histories in each of which Alice determines half of Bob's choice. Now this
becomes a superposition of histories in each of which Alice knows in advance
half of Bob's choice before performing any computation.

\subsection{The mechanism of the speed-up in Grover's algorithm}

We have seen that Gover's algorithm is a superposition of histories in each of
which Alice knows in advance one of the possible halves of Bob's choice. We
note that this holds for any quantum algorithm that reconstructs Bob's choice,
no matter whether with or without speed-up -- the maximally entangled state
(\ref{threed}) is evidently the end state in any case. Thus, at one extreme,
the quantum algorithm can be a superposition of identical histories in each of
which Alice classically reconstructs Bob's choice without benefitting of the
advanced knowledge that tags the history. At the other extreme, in each
history, Alice should be able to perform only the ($\mathcal{N}_{a}$) function
evaluations required to classically reconstruct Bob's choice given the
advanced knowledge of half of it. In fact, this is what is needed to bring the
halved entropy of Bob's register down to zero. This is what Grover's algorithm
does. It goes along with interleaving function evaluations with
non-computational unitary transformations that each time maximize the amount
of information about Bob's choice acquired by Alice with function evaluation.
This minimizes the number of function evaluations required to reconstruct
Bob's choice, bringing it exactly to $\mathcal{N}_{a}$.

We show how things go in detail, starting with the function evaluation part of
the algorithm.

Let us assume that Bob's choice is $\mathbf{b}=01$. Alice's advanced knowledge
can be: $\mathbf{b}\in\left\{  01,00\right\}  $, or $\mathbf{b}\in\left\{
01,11\right\}  $, or $\mathbf{b}\in\left\{  01,10\right\}  $ (Section 2.2). We
assume it is $\mathbf{b}\in\left\{  01,00\right\}  $ (we are pinpointing one
of the possible histories). To identify the value of $\mathbf{b}$ Alice should
compute $\delta\left(  \mathbf{b},\mathbf{a}\right)  $ (for short "$\delta
$")\ for either $\mathbf{a}=01$ or $\mathbf{a}=00$. We assume it is for
$\mathbf{a}=01$. The outcome of the computation, $\delta=1$, tells Alice that
$\mathbf{b}=01$. This corresponds to two classical computation histories, one
for each possible sharp state of register $V$: we represent each classical
computation history as a sequence of sharp quantum states. The initial state
of history 1 is $\operatorname{e}^{i\varphi_{1}}\left\vert 01\right\rangle
_{B}\left\vert 01\right\rangle _{A}\left\vert 0\right\rangle _{V}$, what means
that the input of the computation of $\delta\left(  \mathbf{b},\mathbf{a}%
\right)  $ is $\mathbf{b}=01,~\mathbf{a}=01$; $\left\vert 0\right\rangle _{V}$
is one of the two possible sharp states of register $V$. The state after the
computation of $\delta$ is $\operatorname{e}^{i\varphi_{1}}\left\vert
01\right\rangle _{B}\left\vert 01\right\rangle _{A}\left\vert 1\right\rangle
_{V}$ -- the result of the computation is modulo 2 added to the former content
of $V$. We are using the history amplitudes that reconstruct the quantum
algorithm; our present aim is to show that the quantum algorithm is a
superposition of histories where Alice classically reconstructs Bob's choice
given the advanced knowledge of one of the possible halves of it.

In history 2, the states before/after the computation of $\delta$ are
$-\operatorname{e}^{i\varphi_{1}}\left\vert 01\right\rangle _{B}\left\vert
01\right\rangle _{A}\left\vert 1\right\rangle _{V}\rightarrow-\operatorname{e}%
^{i\varphi_{1}}\left\vert 01\right\rangle _{B}\left\vert 01\right\rangle
_{A}\left\vert 0\right\rangle _{V}$.

In the case that Alice computes $\delta\left(  \mathbf{b},\mathbf{a}\right)  $
for $\mathbf{a}=00$ instead, she obtains $\delta=0$, which of course tells her
again that $\mathbf{b}=01$. This originates other two histories. History 3:
$\operatorname{e}^{i\varphi_{1}}\left\vert 01\right\rangle _{B}\left\vert
00\right\rangle _{A}\left\vert 0\right\rangle _{V}\rightarrow\operatorname{e}%
^{i\varphi_{1}}\left\vert 01\right\rangle _{B}\left\vert 00\right\rangle
_{A}\left\vert 0\right\rangle _{V}$; history 4: $-\operatorname{e}%
^{i\varphi_{1}}\left\vert 01\right\rangle _{B}\left\vert 00\right\rangle
_{A}\left\vert 1\right\rangle _{V}\rightarrow-\operatorname{e}^{i\varphi_{1}%
}\left\vert 01\right\rangle _{B}\left\vert 00\right\rangle _{A}\left\vert
1\right\rangle _{V}$. Etc.

The function evaluation step of Grover's algorithm, namely the transformation
of state (\ref{tred}) into state (\ref{secondstaged}), is the superposition of
all such histories.

Function evaluation is preceded and followed by two non-computational unitary
transformations, respectively $U_{A}U_{V}$ and $U_{A}^{\prime}$. The first
transformation branches the initial sharp state of registers $A$ and $V$ into
the superposition of the inputs of the function evaluation part of the
histories. This superposition maximizes the amount of information acquired by
Alice with function evaluation -- i. e. entanglement between $\hat{B}$ and
$\hat{A}$. The second branches the output states of function evaluation into a
superposition of states that interfere with one another making correlation of
entanglement. As already noted in Section 2.1, these transformations
(together) maximize the correlation between the outcomes of measuring $\hat
{B}$ and $\hat{A}$ respectively at the beginning and the end of the unitary
part of the algorithm. In other words, they maximize the probability of
finding Bob's choice in register $A$.

Summing up, Grover's algorithm for $n=2$ can be decomposed into a
superposition of histories in each of which Alice knows in advance half of the
result of the computation and utilizes this information to identify the other
half in a classical way. This clarifies why, according to the
information-theoretic temporal Bell inequality derived by Morikoshi $\left[
5\right]  $, all is as if Grover's algorithm exploited unperformed
computations. This is what happens in each and every one of the histories
Grover's algorithm is made of.

Let us now consider the case $n>2$. As well known, the sequence "function
evaluation-inversion about the mean" (the algorithm's \textit{iterate}%
)\ should be repeated $\frac{\pi}{4}2^{n/2}$ times. This maximizes the
probability of finding the solution leaving a probability of error $\leq
\frac{1}{2^{n}}$. This goes along with the present explanation of the speed-up
in the order of magnitude. In fact, according to it, one should perform
$\operatorname{O}\left(  2^{n/2}\right)  $ computations of $\delta$ -- this is
the number of classical computations required to find the missing half of
Bob's choice given the advanced knowledge of the other half.

\section{Generalizing the mechanism of the speed-up}

In all the quantum algorithms examined in this paper, finding the solution of
the problem (a deterministic or stochastic function of Bob's choice) is a
by-product of reconstructing Bob's choice. Because of this commonality, all
these algorithms can be generated by a simple generalization of Grover's algorithm.

Given the problem, let $p_{S}$\ be the probability of finding the (or a)
solution with a potential measurement of $\hat{A}$. To generate the quantum
algorithm that solves the problem, we set the matrix elements of the
non-computational transformations of Grover's algorithm free up to unitarity;
then, after the transformation that follows each function evaluation, maximize
$p_{S}$. For the time being we give the generalized algorithm. In the
following sections we will check that it unifies all the quantum algorithms
considered in this paper.

I) Start with some set of functions $f_{\mathbf{b}}\left(  \mathbf{a}\right)
$, with $f$, $\mathbf{b}$, and $\mathbf{a}$ ranging over some sets of values.
For example, in Grover's algorithm, we have $f_{\mathbf{b}}\left(
\mathbf{a}\right)  \equiv\delta\left(  \mathbf{b},\mathbf{a}\right)  $, with
$f$ ranging over $\left\{  0,1\right\}  $ and $\mathbf{b},\mathbf{a}$ over
$\left\{  0,1\right\}  ^{n}$. The imaginary register $B$ contains $\mathbf{b}%
$, the label of the function, register $A$ the argument of the function, and
register $V$ the result of function evaluation reversibly added to its former content.

II) Assume that $B$\ is in a maximally mixed state, prepare $A$ and $V$ in a
sharp state.

III) Apply to $A$ a unitary transformation whose matrix elements are free
variables up to unitarity. Do the same with $V$.

IV) Perform function evaluation.

V) Apply to $A$ another free unitary transformation.

VI) Maximize $p_{S}$, what can be done in principle by zeroing its partial
derivatives with respect to the free variables we are dealing with.

VII) Points (IV), (V), and (VI) constitute the algorithm's\textit{ iterate}.
Iterate until $p_{S}=1$. In all the cases examined, this sets the algorithm to
a superposition of histories in each of which Alice classically reconstructs
Bob's choice given the advanced knowledge of one of the possible halves of it.
The number of function evaluations is always that ($\mathcal{N}_{a}$) foreseen
by the sharing rule.

VIII) Acquire the characteristic by measuring $\hat{A}$.

It should be noted that the present mechanism diverges from Grover's algorithm
if we over-iterate. Having replaced the inversion about the mean by the
unitary transformation that maximizes $p_{S}$, it is never the case that we
reduce this probability -- this transformation becomes the identity if we over-iterate.

A slight modification of this mechanism can be applied to the search for new
speed-ups even if we do not know beforehand which is the characteristic of the
function that leaks to register $A$ with function evaluation (as necessary to
compute  $p_{S}$). Let $\left\vert \psi\right\rangle _{A}$ be the state
(reduced density operator in random phase representation) of register $A$,
$\mathcal{E}_{A}$ its entropy. Clearly, $\mathcal{E}_{A}$ gauges the amount of
information about Bob's choice leaked to register $A$\ with function
evaluation -- for example, it is zero bit in states (\ref{init}) and
(\ref{tred}) and two bits in states (\ref{secondstaged}) and (\ref{threed}).
We should perform steps (I) through (IV) and maximize $\mathcal{E}_{A}$. At
this point, we should try to identify the characteristic of the function
leaked -- what the information leaked is about. For example, this is
relatively simple in Grover's and Deutsch\&Jozsa's algorithms. We note that
this characteristic is fully there, in the part of the state of Alice's
register entangled with Bob's choice, after the first function evaluation.
Eventually, provided that we have succeeded in identifying the characteristic
in question, we can perform steps (V) through (VIII). Reference $\left[
11\right]  $ provides the example of a new speed-up that can be obtained in
this way (finding a certain characteristic of a permutation). Reasonably,
given any set of functions, this mechanism generates with the maximum possible
speed-up a characteristic of the function chosen by Bob. Naturally, we should
look for set of functions where knowing in advance half of Bob's choice yields
an interesting advantage. 

We pinpoint a limit of the result obtained. Maximizing each time the
probability of finding the solution in Alice's register minimizes the number
of function evaluations required to reach it. Whether this number is always
$\mathcal{N}_{a}$ -- the number foreseen by the sharing rule -- is of course
an important question in the present context. For the time being, we must
leave this question open in the general case. This work is exploratory in
character and\ we limit ourselves to checking that the two numbers coincide
with one another in all the quantum algorithms examined.

It might be interesting to underline the kernel of the present mechanism,
which is maximizing in a suitable quantum context input-output correlation.
Quantum retroaction of the output on the input (Alice's action contributing to
Bob's choice) is what allows building this correlation with a speed-up.

\section{Deutsch\&Jozsa's algorithm}

In Deutsch\&Jozsa's $\left[  15\right]  $ algorithm, the set of functions is
all the constant and \textit{balanced} functions (with the same number of
zeroes and ones) $f_{\mathbf{b}}:\left\{  0,1\right\}  ^{n}\rightarrow\left\{
0,1\right\}  $. Array (\ref{dj}) gives (part of) the set of eight\ functions
for $n=2$.%
\begin{equation}%
\begin{tabular}
[c]{cccccc}%
$\mathbf{a}$ & $\,f_{0000}\left(  \mathbf{a}\right)  $ & $f_{1111}\left(
\mathbf{a}\right)  $ & $f_{0011}\left(  \mathbf{a}\right)  $ & $f_{1100}%
\left(  \mathbf{a}\right)  $ & $f_{0101}\left(  \mathbf{a}\right)  $\\\hline
00 & 0 & 1 & 0 & 1 & 0\\\hline
01 & 0 & 1 & 0 & 1 & 1\\\hline
10 & 0 & 1 & 1 & 0 & 0\\\hline
11 & 0 & 1 & 1 & 0 & 1
\end{tabular}
\ \text{etc.}\label{dj}%
\end{equation}

The bit string $\mathbf{b}\equiv b_{0},b_{1},...,b_{2^{n}-1}$ is both the
suffix and the table of the function $f_{\mathbf{b}}\left(  \mathbf{a}\right)
$ -- the sequence of function values for increasing values of the argument.
Specifying the choice of the function by means of the table of the function
simplifies the discussion. Alice is to find whether the function selected by
Bob is balanced or constant by computing $f_{\mathbf{b}}\left(  \mathbf{a}%
\right)  \equiv f\left(  \mathbf{b},\mathbf{a}\right)  $ for appropriate
values of $\mathbf{a}$. In the classical case this requires, in the worst
case, a number of computations of $f\left(  \mathbf{b},\mathbf{a}\right)  $
exponential in $n$; in the quantum case one computation.

\subsection{Time-symmetric representation}

Register $B$\ ($A$)\ contains $\mathbf{b}$\ ($\mathbf{a}$), register $V$\ the
result of function evaluation reversibly added to its former content. The
input and output states of the quantum algorithm are respectively:%
\begin{equation}
\left\vert \psi\right\rangle =\frac{1}{2\sqrt{2}}\left(  \operatorname{e}%
^{i\varphi_{0}}\left\vert 0000\right\rangle _{B}+\operatorname{e}%
^{i\varphi_{1}}\left\vert 1111\right\rangle _{B}+\operatorname{e}%
^{i\varphi_{2}}\left\vert 0011\right\rangle _{B}+\operatorname{e}%
^{i\varphi_{3}}\left\vert 1100\right\rangle _{B}+...\right)  \left\vert
00\right\rangle _{A}\left\vert 1\right\rangle _{V}, \label{twodigei}%
\end{equation}%
\begin{align}
U_{A}U_{f}U_{A}U_{V}\left\vert \psi\right\rangle  &  =\frac{1}{4}\left[
\left(  \operatorname{e}^{i\varphi_{0}}\left\vert 0000\right\rangle
_{B}-\operatorname{e}^{i\varphi_{1}}\left\vert 1111\right\rangle _{B}\right)
\left\vert 00\right\rangle _{A}+\left(  \operatorname{e}^{i\varphi_{2}%
}\left\vert 0011\right\rangle _{B}-\operatorname{e}^{i\varphi_{3}}\left\vert
1100\right\rangle _{B}\right)  \left\vert 10\right\rangle _{A}+...\right]
\nonumber\\
&  \left(  \left\vert 0\right\rangle _{V}-\left\vert 1\right\rangle
_{V}\right)  . \label{threedj}%
\end{align}
$U_{A}$ and $U_{V}$ are the Hadamard transforms on respectively registers $A$
and $V$, $U_{f}$ is function evaluation, namely the computation of $f\left(
\mathbf{b},\mathbf{a}\right)  $. Measuring $\hat{B}$\ in state (\ref{twodigei}%
) yields Bob's choice, a value of $\mathbf{b}$. Measuring $\hat{A}$ in state
(\ref{threedj})\ yields the characteristic of the function: "constant" if
$\mathbf{a}$ is all zeros, "balanced" otherwise.

This time the result of Alice's measurement is not Bob's choice but a function
thereof. However, as we will show in sections 4.3 and 4.4, the determination
of this result is a by-product of reconstructing Bob's choice. This can be
explicitly represented by adding another imaginary register $A^{\prime}$ of
the same size of $B$. Besides reversibly writing in $V$ the result of function
evaluation, the black box should reversibly write in $A^{\prime}$\ the
corresponding reconstruction of Bob's choice. States (\ref{twodigei}) and
(\ref{threedj}) should be replaced respectively by%
\begin{equation}
\frac{1}{2\sqrt{2}}\left(  \operatorname{e}^{i\varphi_{0}}\left\vert
0000\right\rangle _{B}+\operatorname{e}^{i\varphi_{1}}\left\vert
1111\right\rangle _{B}+...\right)  \left\vert 0000\right\rangle _{A^{\prime}%
}\left\vert 00\right\rangle _{A}\left\vert 1\right\rangle _{V} \label{augg}%
\end{equation}

and%
\begin{equation}
\frac{1}{4}\left[  \operatorname{e}^{i\varphi_{0}}(\left\vert
0000\right\rangle _{B}\left\vert 0000\right\rangle _{A^{\prime}}%
-\operatorname{e}^{i\varphi_{1}}\left\vert 1111\right\rangle _{B}\left\vert
1111\right\rangle _{A^{\prime}})\left\vert 00\right\rangle _{A}+...\right]
\left(  \left\vert 0\right\rangle _{V}-\left\vert 1\right\rangle _{V}\right)
. \label{aug2}%
\end{equation}

\subsection{Sharing the determination of Bob's choice}

The determination of Bob's choice should be shared evenly between the
measurements of $\hat{B}$ and $\hat{A}^{\prime}$\ exactly as we did with
$\hat{B}$ and $\hat{A}$ in the case of Grover's algorithm. The fact Alice does
not really measure $\hat{A}^{\prime}$ but a function thereof (i. e. $\hat{A}$)
is irrelevant. The important thing is that Alice would acquire Bob's choice by
measuring $\hat{A}^{\prime}$.

The state of register $B$ in states (\ref{twodigei}) and (\ref{threedj}) -- or
(\ref{augg}) and (\ref{aug2}) -- is:
\begin{equation}
\left\vert \psi\right\rangle _{B}=\frac{1}{2\sqrt{2}}\left(  \operatorname{e}%
^{i\varphi_{0}}\left\vert 0000\right\rangle _{B}+\operatorname{e}%
^{i\varphi_{1}}\left\vert 1111\right\rangle _{B}+\operatorname{e}%
^{i\varphi_{2}}\left\vert 0011\right\rangle _{B}+\operatorname{e}%
^{i\varphi_{3}}\left\vert 1100\right\rangle _{B}+...\right)  . \label{roa}%
\end{equation}

Say that Bob's choice is $\mathbf{b}=0011$. This bit string is the table of
the function chosen by Bob, more explicitly: $f_{\mathbf{b}}\left(  00\right)
=0,f_{\mathbf{b}}\left(  01\right)  =0,f_{\mathbf{b}}\left(  10\right)
=1,f_{\mathbf{b}}\left(  11\right)  =1$. $P_{B}$ is the projection of
$\left\vert \psi\right\rangle _{B}$\ on $\left\vert 0011\right\rangle _{B}$,
namely on the table of the function. We can share $P_{B}$ by taking two shares
of the table such that the projections of $\left\vert \psi\right\rangle _{B}$
on them satisfy the sharing rule (of course such projections can be related to
partial measurements of the content of register $B$). We show further below
that such two shares of the table should be two complementary half tables in
each of which all the values of the function are the same. We call each share
of this kind a \textit{good half table}.

This leaves us with only one way of sharing the table $\mathbf{b}=0011$; the
two shares should be $f_{\mathbf{b}}\left(  00\right)  =0,f_{\mathbf{b}%
}\left(  01\right)  =0$ and respectively $f_{\mathbf{b}}\left(  10\right)
=1,f_{\mathbf{b}}\left(  11\right)  =1$. The former half table corresponds to
the projection of $\left\vert \psi\right\rangle _{B}$ on $\mathbf{b}%
\in\left\{  0011,0000\right\}  $, the latter on $\mathbf{b}\in\left\{
0011,1111\right\}  $. Either half table represents the contribution of Alice's
action to the determination of Bob's choice.

We show that there is no other way of satisfying the sharing rule. First, let
us assume that one of the two complementary half tables is not good (the
values of the function are not all the same). Because of the structure of the
table, also the other half would not be good. Thus, the two corresponding
shares of $P_{B}$\ would both determine the fact that the function is balanced
(a Boolean function of $\mathbf{b}$). This would violate the no
over-determination condition of the sharing rule. If one or both shares were
less than half table, this would either not satisfy equation (\ref{equal}) or
not determine the value of $\mathbf{b}$, as readily checked.

\subsection{Advanced knowledge}

Also in the present case, the fact that Alice contributes to the determination
of Bob's choice implies that she knows that contribution in advance. This can
be seen more quickly as follows. Since the state of register $B$ remains
unaltered throughout the unitary part of Alice's action, also its projection
on the half table remains unaltered. At the end of the unitary part of Alice's
action, this projection represents the contribution of Alice's action to the
determination of Bob's choice. Advanced at the beginning, it changes Alice's
complete ignorance of Bob's choice into knowledge of the half table.

We can see that the quantum algorithm requires the number of function
evaluations of a classical algorithm that has to reconstruct Bob's choice
starting from the advanced knowledge of a good half table. In fact, the value
of $\mathbf{b}$ is always identified by computing $f_{\mathbf{b}}\left(
\mathbf{a}\right)  $ for only one value of $\mathbf{a}$\ (anyone) outside the
half table. Thus, both the quantum algorithm and the advanced knowledge
classical algorithm require just one function evaluation.

\subsection{Mechanism of the speed-up}

Let us group the histories with the same value of $\mathbf{b}$. Starting with
$\mathbf{b}=0011$, we assume that Alice's advanced knowledge is, e. g.,
$\mathbf{b}\in\left\{  0011,0000\right\}  $. In order to determine the value
of $\mathbf{b}$ and thus the characteristic of the function, Alice should
perform function evaluation for either $\mathbf{a}=10$ or $\mathbf{a}=11$. We
assume it is for $\mathbf{a}=10$. Since we are under the assumption
$\mathbf{b}=0011$, the result of the computation is $1$. This, besides telling
Alice that $\mathbf{b}=0011$, originates two classical computation histories,
each consisting of a state before and one after function evaluation. History
1: $\operatorname{e}^{i\varphi_{2}}\left\vert 0011\right\rangle _{B}\left\vert
10\right\rangle _{A}\left\vert 0\right\rangle _{V}\rightarrow\operatorname{e}%
^{i\varphi_{2}}\left\vert 0011\right\rangle _{B}\left\vert 10\right\rangle
_{A}\left\vert 1\right\rangle _{V}$. History 2: $-\operatorname{e}%
^{i\varphi_{2}}\left\vert 0011\right\rangle _{B}\left\vert 10\right\rangle
_{A}\left\vert 1\right\rangle _{V}\rightarrow-\operatorname{e}^{i\varphi_{2}%
}\left\vert 0011\right\rangle _{B}\left\vert 10\right\rangle _{A}\left\vert
0\right\rangle _{V}$.\ If she performs function evaluation for $\mathbf{a}=11$
instead, this originates other two histories, etc.

As readily checked, the superposition of all these histories is the function
evaluation stage of the quantum algorithm. Then, Alice applies the Hadamard
transform to register $A$. Each history branches into four histories. The end
states of such branches interfere with one another to yield state
(\ref{threedj}).

We can see that Deutsch\&Jozsa algorithm is generated by the mechanism of the
speed-up of Section 3. We should replace the Hadamard transforms before and
after function evaluation by free unitary transformations and then maximize
$p_{S}$ (the probability of finding the solution in register $A$).

It is easy to see that the present analysis, like the notion of sharing the
table of the function into two complementary good halves, holds unaltered for
$n>2$.

\section{Simon's and the hidden subgroup algorithms}

In Simon's $\left[  16\right]  $ algorithm, the set of functions is all the
$f_{\mathbf{b}}:\left\{  0,1\right\}  ^{n}\rightarrow\left\{  0,1\right\}
^{n-1}$ such that $f_{\mathbf{b}}\left(  \mathbf{a}\right)  =f_{\mathbf{b}%
}\left(  \mathbf{c}\right)  $ if and only if $\mathbf{a}=\mathbf{c}$\ or
$\mathbf{a}=\mathbf{c}\oplus\mathbf{h}^{\left(  \mathbf{b}\right)  }$;
$\oplus$\ denotes bitwise modulo 2 addition; the bit string $\mathbf{h}%
^{\left(  \mathbf{b}\right)  }$\textbf{, }depending on $\mathbf{b}$ and
belonging to $\left\{  0,1\right\}  ^{n}$ excluded the all zeroes string, is a
sort of period of the function. Array (\ref{periodic}) gives (part of) the set
of six functions for $n=2$. The bit string $\mathbf{b}$ is both the suffix and
the table of the function. Since $\mathbf{h}^{\left(  \mathbf{b}\right)
}\oplus\mathbf{h}^{\left(  \mathbf{b}\right)  }=\mathbf{0}$ (the all zeros
string), each value of the function appears exactly twice in the table, thus
50\% of the rows plus one always identify $\mathbf{h}^{\left(  \mathbf{b}%
\right)  }$.
\begin{equation}%
\begin{tabular}
[c]{ccccc}\hline
& $\mathbf{h}^{\left(  0011\right)  }=01$ & $\mathbf{h}^{\left(  1100\right)
}=01$ & $\mathbf{h}^{\left(  0101\right)  }=10$ & $\mathbf{h}^{\left(
1010\right)  }=10$\\\hline
$\mathbf{a}$ & $f_{0011}\left(  \mathbf{a}\right)  $ & $f_{1100}\left(
\mathbf{a}\right)  $ & $f_{0101}\left(  \mathbf{a}\right)  $ & $f_{1010}%
\left(  \mathbf{a}\right)  $\\\hline
00 & 0 & 1 & 0 & 1\\\hline
01 & 0 & 1 & 1 & 0\\\hline
10 & 1 & 0 & 0 & 1\\\hline
11 & 1 & 0 & 1 & 0
\end{tabular}
\ \text{etc.}\label{periodic}%
\end{equation}

Bob selects a value of $\mathbf{b}$. Alice's problem is finding the value of
$\mathbf{h}^{\left(  \mathbf{b}\right)  }$, "hidden" in $f_{\mathbf{b}}\left(
\mathbf{a}\right)  $, by computing $f_{\mathbf{b}}\left(  \mathbf{a}\right)
=f\left(  \mathbf{b},\mathbf{a}\right)  $ for different values of $\mathbf{a}%
$. In present knowledge, a classical algorithm requires a number of
computations of $f\left(  \mathbf{b},\mathbf{a}\right)  $ exponential in $n$.
The quantum algorithm solves the hard part of this problem, namely finding a
string $\mathbf{s}_{j}^{\left(  \mathbf{b}\right)  }$ orthogonal to
$\mathbf{h}^{\left(  \mathbf{b}\right)  }$, with one computation of $f\left(
\mathbf{b},\mathbf{a}\right)  $; "orthogonal" means that the modulo 2 addition
of the bits of the bitwise product of the two strings is zero. There are
$2^{n-1}$ such strings. Running the quantum algorithm yields one of these
strings at random (see further below). The quantum algorithm is iterated until
finding $n-1$ different strings. This allows us to find $\mathbf{h}^{\left(
\mathbf{b}\right)  }$ by solving a system of modulo 2 linear equations.

We check that the history superposition picture and the mechanism of the
speed-up for the present algorithm.

\subsection{Time-symmetric representation}

Register $B$\ ($A$)\ contains $\mathbf{b}$\ ($\mathbf{a}$), register $V$\ the
result of function evaluation reversibly added to it former content. The input
and output states of the quantum algorithm are respectively:%

\begin{equation}
\left\vert \psi\right\rangle =\frac{1}{\sqrt{6}}\left(  \operatorname{e}%
^{i\varphi_{0}}\left\vert 0011\right\rangle _{B}+\operatorname{e}%
^{i\varphi_{1}}\left\vert 1100\right\rangle _{B}+\operatorname{e}%
^{i\varphi_{2}}\left\vert 0101\right\rangle _{B}+\operatorname{e}%
^{i\varphi_{3}}\left\vert 1010\right\rangle _{B}+...\right)  \left\vert
00\right\rangle _{A}\left\vert 0\right\rangle _{V}, \label{dues}%
\end{equation}%
\[
U_{A}U_{f}U_{A}\left\vert \psi\right\rangle =
\]%
\begin{equation}
\frac{1}{2\sqrt{6}}\left\{
\begin{array}
[c]{c}%
(\operatorname{e}^{i\varphi_{0}}\left\vert 0011\right\rangle _{B}%
+\operatorname{e}^{i\varphi_{1}}\left\vert 1100\right\rangle _{B})\left[
(\left\vert 00\right\rangle _{A}+\left\vert 10\right\rangle _{A})\left\vert
0\right\rangle _{V}+(\left\vert 00\right\rangle _{A}-\left\vert
10\right\rangle _{A})\left\vert 1\right\rangle _{V}\right]  +\\
(\operatorname{e}^{i\varphi_{2}}\left\vert 0101\right\rangle _{B}%
+\operatorname{e}^{i\varphi_{3}}\left\vert 1010\right\rangle _{B})\left[
(\left\vert 00\right\rangle _{A}+\left\vert 01\right\rangle _{A})\left\vert
0\right\rangle _{V}+(\left\vert 00\right\rangle _{A}-\left\vert
01\right\rangle _{A})\left\vert 1\right\rangle _{V}\right]  +...
\end{array}
\right\}  . \label{tres}%
\end{equation}

In state (\ref{dues}), $V$ is prepared in the all zeros string (just one zero
for $n=2$). $U_{A}$ is Hadamard on $A$, $U_{V}$ -- being the identity here --
does not appear, $U_{f}$ is function evaluation. In state (\ref{tres}), for
each value of $\mathbf{b}$, register $A$ (no matter the content of $V$) hosts
even weighted\ superpositions of the $2^{n-1}$ strings $\mathbf{s}%
_{j}^{\left(  \mathbf{b}\right)  }$ orthogonal to $\mathbf{h}^{\left(
\mathbf{b}\right)  }$. By measuring $\hat{A}$ in this state, Alice obtains at
random one of these $\mathbf{s}_{j}^{\left(  \mathbf{b}\right)  }$. Then she
repeats the "right part" of the algorithm (preparation of registers $A$\ and
$V$, computation of $f\left(  \mathbf{b},\mathbf{a}\right)  $, and measurement
of $\hat{A}$) until obtaining $n-1$ different $\mathbf{s}_{j}^{\left(
\mathbf{b}\right)  }$.

As we will see in sections 5.3 and 5.4, finding the characteristic of the
function is a by-product of reconstructing Bob's choice. \ We omit the
explicit representation of this reconstruction, completely similar to that of
Section 4.

\subsection{Sharing the determination of Bob's choice}

This time a good half table should not contain a same value of the function
twice, what would over-determine $\mathbf{h}^{\left(  \mathbf{b}\right)  }$,
namely a Boolean function of $\mathbf{b}$ (also the other half would contain a
same value twice). Assume Bob's choice is $\mathbf{b}=0011$. There are two
ways of sharing this table. One is $f_{\mathbf{b}}\left(  00\right)
=0,f_{\mathbf{b}}\left(  10\right)  =1$ and $f_{\mathbf{b}}\left(  01\right)
=0,f_{\mathbf{b}}\left(  11\right)  =1$; the corresponding shares of $P_{B}%
$\ are the projections of $\left\vert \psi\right\rangle _{B}$ on
$\mathbf{b}\in\left\{  0011,0110\right\}  $ and $\mathbf{b}\in\left\{
0011,1001\right\}  $. The other is $f_{\mathbf{b}}\left(  00\right)
=0,f_{\mathbf{b}}\left(  11\right)  =1$ and $f_{\mathbf{b}}\left(  01\right)
=0,f_{\mathbf{b}}\left(  10\right)  =1$, etc.

We should note that sharing each table into two halves is accidental to the
present algorithm. In the quantum part of Shor's $\left[  17\right]  $
factorization algorithm (finding the period of a periodic function), taking
two parts of the table that do not contain a same value of the function twice
implies that each part is less than half table if the domain of the function
spans more than two periods.

\subsection{Advanced knowledge}

Ascribing to Alice's action the determination of a good half table implies
that she knows it in advance -- as in Section 4.3. Also in the present case
the quantum algorithm requires the number of function evaluations of a
classical algorithm that has to determine Bob's choice starting from the
advanced knowledge of a good half table. In fact, since no value of the
function appears twice in the half table, the value of $\mathbf{b}$ is always
identified by computing $f\left(  \mathbf{b},\mathbf{a}\right)  $ for only one
value of $\mathbf{a}$\ (anyone) outside the half table.

\subsection{Mechanism of the speed-up}

The history superposition picture can be developed as in Section 4.4: given
the advanced knowledge of, say, $\mathbf{b}\in\left\{  0011,0110\right\}  $,
in order to determine the value of $\mathbf{b}$, Alice should perform function
evaluation for either $\mathbf{a}=01$ or $\mathbf{a}=11$, etc. We can see that
Simon's algorithm is generated by the mechanism of the speed-up of Section 3
(here the solution, any $\mathbf{s}_{j}^{\left(  \mathbf{b}\right)  }$
orthogonal to $\mathbf{h}^{\left(  \mathbf{b}\right)  }$, is stochastic in
character). We should replace the transformations before and after function
evaluation (comprising the identity on register $V$) by free unitary
transformations and then maximize $p_{S}$ (the probability of finding the
solution in register $A$).

The present analysis -- like the notion of sharing the table into two good
halves -- holds unaltered for $n>2$. It also applies to the generalized
Simon's problem and to the Abelian hidden subgroup problem. In fact the
corresponding algorithms are essentially the same as the algorithm that solves
Simon's problem. In the hidden subgroup problem, the set of functions
$f_{\mathbf{b}}:G\rightarrow W$ map a group $G$ to some finite set $W$\ with
the property that there exists some subgroup $S\leq G$ such that for any
$\mathbf{a},\mathbf{c}\in G$, $f_{\mathbf{b}}\left(  \mathbf{a}\right)
=f_{\mathbf{b}}\left(  \mathbf{c}\right)  $ if and only if $\mathbf{a}%
+S=\mathbf{c}+S$. The problem is to find the hidden subgroup $S$ by computing
$f_{\mathbf{b}}\left(  \mathbf{a}\right)  $ for various values of $\mathbf{a}%
$. Now, a large variety of problems solvable with a quantum speed-up can be
re-formulated in terms of the hidden subgroup problem $\left[  18,19\right]
$. Among these we find: the seminal Deutsch's problem, finding orders, finding
the period of a function (thus the problem solved by the quantum part of
Shor's\ factorization algorithm), discrete logarithms in any group, hidden
linear functions, self shift equivalent polynomials, Abelian stabilizer
problem, graph automorphism problem.

\section{Discussion and conclusions}

We have pinpointed the fundamental reason for which quantum algorithms can
require fewer function evaluations than the minimum required by any equivalent
classical algorithm and/or violate Morikoshi's information-theoretic temporal
Bell inequality. The quantum principle, stating that the measurement of an
observable determines one of its eigenvalues, becomes ambiguous when the
measurement of two commuting observables yields at random two identical
eigenvalues, which in our case are Bob's choice and Alice's reconstruction of
it. Which measurement determines their common value? Postulating that the
projection of the quantum state induced by either measurement (i. e. the
determination) shares between the two measurements (i) with no
over-projection, (ii) with entropy reductions the same for each share, and
(iii) in a uniform quantum superposition of all the possible ways of sharing
compatible with the former conditions, implies that the quantum algorithm is a
uniform superposition of algorithms (histories) in each of which Alice
determines one of the possible halves of Bob's choice. Advancing this
determination to the beginning of Alice's action shows that Alice, in each
history, knows in advance half of Bob's choice. In all the cases examined, she
can perform only the $\mathcal{N}_{a}$ function evaluations required to
classically reconstruct Bob's choice given the advanced knowledge of half of it.

To this end, function evaluations should be interleaved with non-computational
unitary transformations that each time maximize the probability of finding the
solution in Alice's register. This also maximizes the amount of information
about Bob's choice acquired by Alice with function evaluation. The number of
function evaluations is correspondingly minimized and brought in fact to
$\mathcal{N}_{a}$ in all the cases examined.

We discuss these results.

The history superposition picture highlights an essential difference between
quantum and classical causality. The former can host a loop of the latter. The
causal quantum process is for example the unitary transformation of
$\left\vert \Psi\right\rangle =\left\vert 01\right\rangle _{B}\left\vert
00\right\rangle _{A}\left\vert 1\right\rangle _{V}$\ into $U_{A}^{\prime}%
U_{f}U_{A}U_{V}\left\vert \Psi\right\rangle =\frac{1}{\sqrt{2}}\left\vert
01\right\rangle _{B}\left\vert 01\right\rangle _{A}\left(  \left\vert
0\right\rangle _{V}-\left\vert 1\right\rangle _{V}\right)  $ -- equations
(\ref{initb}) and (\ref{quattrod}). This is a superposition of histories in
each of which Alice knows in advance half of the result of her computation and
exploits this information to reach that same result with fewer function
evaluations. Alice's partial knowledge of the result of a computation before
performing it (a causality loop and in fact the reason for the violation of
Morikoshi's inequality)\ would be impossible if histories were isolated with
respect to one another. However, quantum superposition and interference (as
generated by the maximization procedure) allow this. The half choice known in
advance in one history becomes the missing half in another one, where it is
computed. Thus, all the possible halves of Bob's choice are computed, in
quantum superposition. Moreover, histories are not isolated from one another,
as quantum interference provides cross-talk between them.

It is natural to think that such loops of classical causality, besides the
violation of Morikoshi's inequality in the case of Grover's algorithm, explain
the violation of temporal Bell inequalities on the part of quantum mechanics.
A way of investigating this prospect is trying and extend the present
explanation of the speed-up to more general quantum processes that yield a
speed-up, like for example quantum random walks $\left[  20\right]  $ or mixed
state quantum computing $\left[  21\right]  $. In a way, we should go back to
the original Feynman's observation that the classical simulation of a quantum
process can require an essentially higher amount of resources $\left[
22\right]  $. As it is, the explanation requires seeing a problem in the input
of the quantum process and the solution of the problem in the output. To apply
it to more general quantum processes, we should decouple it from
problem-solving. This would seem to be possible. The basic concept of the
explanation is the possibility that the quantum process builds a stronger than
classical input-output correlation thanks to the fact that (from the
standpoint of quantum correlation) the final measurement of the output
contributes to determining the input. This concept of quantum retroaction of
the output on the input is not committed to problem-solving.

From a technical standpoint, the present work can be used in the search for
new speed-ups. Given a set of functions, one should: (a) interleave function
evaluations with free non-computational unitary transformations, (b) after the
unitary transformation that follows the first function evaluation, maximize
the amount of information about Bob's choice leaked to Alice's register, (c)
identify the characteristic of the function obtained and (d) iterate function
evaluation and the successive unitary transformation maximizing each time the
probability of finding that characteristic in Alice's register. The number of
function evaluations should be that required to reconstruct Bob's choice given
the advanced knowledge of half of it according to the sharing rule. Reference
$\left[  11\right]  $ provides the example of a new quantum speed-up that can
be obtained in this way.

In conclusion, although preliminary in character, these results seem to open a
gap in a problem that has remained little explored. Until now there was no
fundamental explanation of the speed-up, no general mechanism for producing it.

\subsection*{Acknowledgments}

Thanks are due to Vint Cerf, David Deutsch, Artur Ekert, Avshalom Elitzur,
David Finkelstein, Hartmut Neven, and Daniel Sheehan for useful comments/discussions.

\subsection*{References}

$\ \ \ \left[  1\right]  $\ Grover L K 1966\textit{ Proc. of the 28th Annual
ACM Symposium on the Theory of Computing, May 22-24} ACM press New York p. 219

$\left[  2\right]  $ Deutsch D 1985 \textit{Proc. Roy. Soc. London} A
\textbf{400} 97

$\left[  3\right]  $ Morikoshi F 2006 Phys. Rev. A \textbf{73} 052308

$\left[  4\right]  $ Leggett A J and Garg A 1985 \textit{Phys. Rev. Lett.}
\textbf{54} 857

$\left[  5\right]  $ Braunstein S L and Caves C M 1988 \textit{Phys. Rev.
Lett.} \textbf{61} 662

$\left[  6\right]  $ Morikoshi F 2011\textit{ Int. J. Theor. Phys.}
\textbf{50} 1858

$\left[  7\right]  $\ Aharonov Y, Bergmann P G and Lebowitz J L 1964
\textit{Phys. Rev. }B \textbf{134 }1410

$\left[  8\right]  $ Vaidman L 2009 Compendium of Quantum Physics: Concepts,
Experiments, History and Philosophy \textit{Greenberger D, Hentschel K and
Weinert F, eds. Springer-Verlag, Berlin Heidelberg}

$\left[  9\right]  $\ Rovelli C 1996 \textit{Int. J. Theor. Phys.} \textbf{35}
1637$\ $

$\left[  10\right]  $\ Dolev S and Elitzur A C 2001 Non-sequential behavior of
the wave function, arXiv:quant-ph/0102109v1

$\left[  11\right]  $\ Castagnoli G 2010 \textit{Phys. Rev. }A \textbf{82} 052334

$\left[  12\right]  $ Castagnoli G 2011 \textit{Proc. of the 92}$^{nd}%
$\textit{\ Annual Meeting of the AAAS Pacific Division, Quantum
Retrocausation: Theory and Experiment, San Diego}$\ $

$\left[  13\right]  $ Bohm D and Pines D 1953 \textit{Phys. Rev}. \textbf{92} 609

$\left[  14\right]  $ Hawking S 2003 On the Shoulders of Giants
\textit{Running Press, Philadelphia-London} p. 731

$\left[  15\right]  $\ Deutsch D and Jozsa R 1992 \textit{Proc. R. Soc. London
}A \textbf{439} 553

$\left[  16\right]  $ Simon D 1994 \textit{Proc. of the 35th Annual IEEE
Symposium on the Foundations of Computer Science} p. 116

$\left[  17\right]  $ Shor P W 1994 \textit{Proc. of the 35th Annual IEEE
Symposium on the Foundations of Computer Science }p. 124

$\left[  18\right]  $ Mosca M and Ekert A 1999\textit{ Lecture Notes in
Computer Science} \textbf{1509}

$\left[  19\right]  $ Kaye P, Laflamme R and Mosca M 2007 An introduction to
Quantum Computing \textit{Oxford University Press} p. 146

$\left[  20\right]  $ Childs A M, Cleve, Deotto E, Farhi E, Gutmann S and
Spielman D A 2003 \textit{Proc.35th ACM Symposium on Theory of Computing} p. 59

$\left[  21\right]  $ Knill E and Laflamme R 1998 \textit{Phys. Rev. Lett.}
\textbf{81} 5672

$\left[  22\right]  $ Feynman R 1982 \textit{Int. J. Theor. Phys.} \textbf{21} 467

\end{document}